\documentclass[twocolumn,astrosymb]{aastex7}
\usepackage{amsmath}
\shorttitle{Radiating Bondi Flows}
\shortauthors{Bailey et al.}
\received{\today}
\revised{\today}
\accepted{\today}
\submitjournal{MNRAS}
\begin{document}

\title{Radiating Bondi Flows II: Giant Planet Accretion Models}

\author[orcid=0000-0002-6940-3161, gname=Avery, sname=Bailey]{Avery P. Bailey}
\affiliation{Steward Observatory, University of Arizona, 933 North Cherry Avenue, Tucson, AZ 85721-0065, USA}
\email[show]{averybailey@arizona.edu}

\author[0000-0001-5253-1338, gname=Kaitlin, sname=Kratter]{Kaitlin M. Kratter}
\affiliation{Steward Observatory, University of Arizona, 933 North Cherry Avenue, Tucson, AZ 85721-0065, USA}
\email{kkratter@arizona.edu}

\author[0000-0002-3644-8726, gname=Andrew, sname=Youdin]{Andrew N.\ Youdin}
\affiliation{Steward Observatory, University of Arizona, 933 North Cherry Avenue, Tucson, AZ 85721-0065, USA}
\email{youdin@arizona.edu}

\begin{abstract}
In the core accretion model of giant planet formation, the late stages of runaway growth are regulated by the hydrodynamic infall of gas from the protoplanetary disk. For a subset of planet-disk pairings, this scenario is analogous to the classical Bondi problem, which has motivated a Bondi-like parameterization of accretion in some population synthesis models. Existing models and the associated classical Bondi rate however, are predicated upon an adiabatic equation of state. In reality, the planet and its associated accretion shock supply a luminosity that substantially heats the accretion flow. In Paper I of this series, we demonstrate that such radiative feedback can dramatically suppress accretion by orders of magnitude. Here we quantify this effect under realistic planet-forming conditions. We find that for planets forming in an unperturbed disk, accretion is suppressed by 1-2 orders of magnitude interior to $\sim 10$ AU. For planets that open a gap, this feedback is less dramatic and the effect is $\sim$ 1 order in magnitude. We investigate the effect of various assumptions regarding dust opacities, shock efficiency, and planet radius and find this radiative suppression mechanism to be fairly insensitive to these effects. We also perform full time-dependent simulations demonstrating that the associated adverse entropy profiles are accurate and stable to convection. A simple and flexible set of open-source tools are provided to incorporate this radiative feedback into existing accretion models and population synthesis frameworks. 

\end{abstract}



\section{Introduction}
In the core accretion paradigm, the later stages of giant planet formation are marked by a period of rapid runaway accretion. When a planet's gaseous envelope becomes approximately equal in mass to its solid core, adding mass to the envelope also has the effect of shortening the envelope cooling time, accelerating the addition of further mass in a compounding process and marking the start of the runaway phase \citep{BodenheimerPollack1986, Pollack+1996}. The accretion rate continues to increase until reaching a magnitude the protoplanetary disk cannot sustain, at which point the planetary envelope contracts \citep{GinzburgChiang2019-contraction}. The accretion rate in this runaway regime and afterwards is limited in combination by viscous transport in the protoplanetary disk, multi-dimensional effects \citep{TanigawaWatanabe2002, LubowDangelo2006}, and a maximal planetary accretion rate. This maximal planetary accretion rate is either estimated from fits to multi-dimensional numerical simulations \citep{TanigawaWatanabe2002, Machida+2010} e.g. \citet{Bitsch+2015, Brugger+2018, KimuraIkoma2022}, or modeled as a Bondi-like accretion rate.  While application of Bondi accretion is most appropriate for lower mass (sub-thermal) planets or polar flows, where infall is expected to be more radial in nature, similar parameterizations have been invoked for higher masses, where rotational effects and vertical stratification become non-negligible \citep{Choksi+2023}.

As such, Bondi accretion often finds its way into models of these later stages of giant planet formation. Planetary population synthesis models, used to produce artificial populations for comparison to the observed exoplanet sample, will sometimes employ a Bondi parameterization e.g.\ the Bern models \citep{Mordasini+2012, Emsenhuber+2021}. Spectral energy distribution calculations \citep{ChoksiChiang2024} of protoplanets invoke Bondi envelope profiles for optically thicker regimes where planet envelopes are found to be more spherical \citep{Fung+2019, Krapp+2024}. Evolutionary models of planets at large orbital separations are also based upon Bondi prescriptions and extended to include other processes like gap opening \citep{GinzburgChiang2019-gap} or make inferences about the properties of protoplanet candidates proposed to sculpt observed protoplanetary disk substructures \citep{GinzburgChiang2019-contraction}.

All these prescriptions however, are based upon the original and most simple \citep{Bondi1952} model, in which a planet's accretion rate depends only on planet mass, the environmental sound speed, and weakly upon the adiabatic index $\gamma$. This simplicity is, in part, a result of the simple thermodynamics in which isentropy is assumed. In reality, the accretion flow is subject to heating/cooling via its own radiation and that of the young luminous planet. Heating or cooling can modify the entropy of the flow thereby changing the sound speed, the location of the sonic point, and the overall steady-state accretion rate. Through this mechanism, one expects the steady-state Bondi accretion rate may also depend on properties like the luminosity and optics of the accretion flow, complicating the simple Bondi rate and leading to possibly diverse planet outcomes. Thus, self-consistent treatment of this radiative feedback and its effect on the accretion rate, which is entirely absent from existing studies, is the focus of this work.

In the preceding work (Paper I), it was shown that inclusion of this radiative feedback mechanism tends to suppress accretion under conditions of high luminosity, optical depth, and slow cooling. This was demonstrated via an idealized set of 1D steady-state models employing constant opacities. The role of this work is to extend those simple yet informative models to realistic planet-forming conditions and quantify the magnitude and robustness of this suppression mechanism. We therefore focus on the computation of $f_{\rm acc}\equiv \dot{M}/\dot{M}_{\rm ad}$, which is a correction factor for a planet's steady-state accretion rate $\dot{M}$ relative to that same planet's standard adiabatic Bondi rate $\dot{M}_{\rm ad}$. $f_{\rm acc}$ is primarily a function of protoplanetary disk properties, planet mass, and semi major axis. We choose this parameterization to streamline adoption of these corrections within existing computations that use standard Bondi rates. We develop tools\footnote{https://github.com/apbailey/radiative-bondi-products\label{repo}} necessary to implement $f_{\rm acc}$ into existing models in a flexible and simple manner (see Section \ref{sec:tools}).

The structure of this paper is as follows. In Section \ref{sec:methods}, we briefly restate the steady-state equations (Paper I) and setup for time-dependent simulations. Section \ref{sec:models} presents a fiducial set of steady-state solutions for realistic protoplanetary disk conditions and corresponding maps of $f_{\rm acc}$. In Section \ref{sec:convect}, we present full time-dependent simulations to confirm the robustness of the solutions presented in \ref{sec:models} -- in particular, their stability to convection. In Section \ref{sec:tools}, we describe our open-source tools\footref{repo} for population synthesis codes to incorporate the effects of radiative feedback. Finally, we summarize our findings in Section \ref{sec:end}. 

\section{Equations and Methods}\label{sec:methods}
\subsection{Steady-State Initial Value Problem}\label{sec:ivp}
To determine the 1D radiative steady-state accretion flow onto a planet in Section \ref{sec:models}, we solve an initial value problem in radius for five dependent variables: density $\rho$, Mach number $\mathcal{M}\equiv v/c_s$, entropy\footnote{this is technically a dimensionless entropy obtained by dividing the true entropy by the specific heat at constant volume $S/c_V$} $s$, luminosity $L$, and radiation energy density $E_r$. The equations solved are as follows (see Paper I for details):
\begin{subequations}\label{eq:ivp}
\begin{align}\label{eq:rho}
&\partial_r \ln\rho = \frac{2}{\gamma+1} \left(-\frac{2}{r} - \partial_r \ln \mathcal{M} - \frac{\partial_r s}{2}\right)\\
&\begin{aligned}\label{eq:mom}
\partial_r \ln \mathcal{M}  =& \frac{(\gamma+1)( 2c_s^2r - GM - c_s^2r^2\partial_r s/\gamma)}{2c_s^2 r^2\left(\mathcal{M}^2-1\right)}\\
& + \frac{\gamma-1}{r} -  \frac{\partial_r s}{2}
\end{aligned}\\\label{eq:s}
%&\partial_r s = \frac{\gamma(\gamma-1)}{\rho c_s^3\mathcal{M}}\frac{\partial_r L}{4\pi r^2}\\
%ANY You can switch back, I think it's cleaner to keep Mdot explicit
&\partial_r s = \frac{\gamma(\gamma-1)}{\dot{M} c_s^2}{\partial_r L}\\
& \partial_r L = 4\pi r^2\rho\kappa_P c(a_r T^4 - E_r)\\
&\partial_r E_r = -\frac{L}{4\pi r^2 c}\left(\frac{2}{r} + \frac{2\partial_r L}{L} + 3\rho \kappa_R\right)\label{eq:er}.
\end{align}
\end{subequations}
with $\gamma$ the ratio of specific heats, $M$ the planet mass, $a_r$ the radiation constant, $c$ the speed of light, and $\kappa_P$, $\kappa_R$ the Planck and Rosseland mean opacities respectively. The system is closed by an ideal gas equation of state relating the temperature $T$ to the sound speed $c_s$,
\begin{equation}
T = \frac{\mu m_pc_s^2}{\gamma k_B}
\end{equation}
with the mean molecular weight $\mu$, proton mass $m_p$, and Boltzmann constant $k_B$. 

In Paper I, the outer boundary condition was taken to be at arbitrarily large radius, i.e.\ $ (\rho_\infty, s_\infty, L_\infty, E_{r,\infty})$, with the condition on $\mathcal{M}$ set by requiring the accretion solution to pass through a sonic point. In the planetary context, it is not so appropriate to integrate from arbitrarily large radius as protoplanetary disk has a finite extent set by the large scale physics. For this application then, we begin integrations from a boundary condition at a disk-scale height $r=H$ where thermodynamic quantities like $\rho_\infty$, $T_\infty$ are well-defined and of the same order as the midplane values. We retain the $\infty$ subscripts to denote any boundary values for consistency with Paper I, but note that the physical equivalence is not one-to-one. Like Paper I, we take the flow to be in radiative equilibrium at the outer boundary $E_{r,\infty} = a_r T_\infty^4$ and $L_\infty$ to be a free parameter (at least for the purposes of computing individual solutions).

Given values for $(M, \rho_\infty, T_\infty,\gamma,\mu, L_\infty, \kappa_R, \kappa_P)$ an accretion model for a given planet is well-defined and may be integrated to recover a mass accretion rate according to the solution method constructed in Paper I. For all models in this work we take $\mu=2.4$, $\gamma=7/5$ for a molecular hydrogen-helium mixture. While $\gamma=7/5$ is not expected everywhere (particularly when rotational modes freeze-out in outer disk), changes to $\gamma$ are expected to cause order unity corrections, so we take this intermediate value as representative. $\kappa_R$ and $\kappa_P$ are taken to be density and temperature dependent functions computed from the opacity tables of \citet{Zhu+2021} assuming solar metallicity and a dust-to-gas ratio of $1/100$. Thus a particular steady-state accretion model in this work may be uniquely defined by the choices of ($M, \rho_\infty$, $T_\infty$, $L_\infty$). 

\subsubsection{Athena++ Simulations}\label{sec:time}
As an alternative to solving the steady-state equations outright, it is orders of magnitude more expensive but still computationally feasible to arrive at a steady-state solution by evolving the full set of time-dependent radiation-hydrodynamics equations for a sufficiently long time. To confirm the results of our steady-state integrations and test their robustness to potential instability, we also run 1D and 2D radiation-hydrodynamics simulations to steady-state with Athena++ \citep{Stone+2020}. As an added benefit, time-dependent simulations are not plagued by any difficulties associated with the sonic point and thus the solution interior to the sonic point may also be simulated and verified. 

We use the frequency-integrated implicit radiation scheme of \citet{Jiang2021} and refer the reader to equation (9) of that work for a summary of the equations solved. The steady-state radiation system in Section \ref{sec:ivp} was formulated with the same frequency-integrated treatment in \citet{Jiang2021} (i.e. the same use of $\kappa_P$, $\kappa_R$), but contains several differences to make solution more tractable. For one, in the steady-state equations, the radiative transfer is taken to the lowest order in $v/c$ -- the radiation sees no distinction between the comoving fluid frame and an inertial lab frame. The effect is small enough in the planet-forming regimes studied here that this is not expected to make a difference. A more substantial, but still potentially small, difference is the closure relation implicit in our steady-state equations relating radiation energy density, flux, and pressure ($E_r = 3P_r - 2F_r/c$). While this closure is chosen to retain appropriate optically thin and thick limits \citep{Shu1992}, there is no guarantee to its accuracy, particularly in the intermediate regimes common to planet-forming environments. The simulations, which solve the radiative transfer equation outright, provide a test of this simplifying assumption in the steady-state formulation. 

In the Athena++ framework, boundary conditions are specified by the setting of density $\rho$, velocity $v$, pressure $p$, and intensity $I(\hat{n})$ along direction $\hat{n}$, in ghost cells. When reproducing our steady-state model in Athena++, we place the outer boundary at $r=H$ and make it so the thermodynamic state of ghost cell variables $(\rho,p)$ there are the same as gas with density, entropy $(\rho_\infty, s_\infty)$. Formally, we set the velocity in the outer ghost cells to $v=0$. But because the boundary flux is determined by the evaluation of a Riemann problem between the last active cell and the adjacent ghost cell, this choice does not amount to zero mass flux at the boundary. Finally, at the outer boundary, we set isotropic inward intensity $I = cE_{r,\infty}/4\pi = acT_\infty^4/4\pi$ set by the disk equilibrium temperature. Near the planetary surface, we place an inner boundary with inflow conditions on the hydrodynamic variables ($\rho$, $v$, $p$ in the ghost cells are held equal to the values in the first active cell). We fix isotropic intensity $I\approx L_{\rm src}/4\pi^2 r_{\rm in}^2$ for intensities along rays which point into the domain ($\theta <\pi/2$) at the inner boundary. For intensities which point out of the domain ($\theta > \pi/2$), we set $I=0$. This results in a flux $F_r = 2\pi \int_0^{\pi/2} I\cos\theta \sin\theta d\theta = \pi I \approx L_{\rm src}/4\pi r^2$. The value for the luminosity at the inner boundary $L_{\rm src}$ is set by hand from steady-state solutions of Equations (\ref{eq:rho})-(\ref{eq:er}) that the simulations are being compared against.

The remaining details of each simulation including resolution, fiducial parameters, and various simulation specific subtleties are relegated to Section \ref{sec:convect}, where results of these simulations are also presented.

\section{Steady-State Models and Accretion Rates}\label{sec:models}
\subsection{Fiducial Parameters}\label{sec:fiducial}
Here we present our 1D steady-state radiative models and corresponding accretion rates. We neglect self-gravity of the accretion flow which is generally reasonable except for the least massive planets at large orbital radii (see Appendix \ref{sec:selfg}). As stated in Section \ref{sec:ivp}, an accretion model for a given planet in this work is defined by specifying four free parameters ($M, \rho_\infty$, $T_\infty$, $L_\infty$). We simplify the problem further however, by defining a fiducial protoplanetary disk model with temperature and surface density profiles in terms of stellocentric coordinate $R$
\begin{equation}
    T_\infty(R) = 300 \text{ K}\left(\frac{R}{1\text { AU}}\right)^{-1/2}
\end{equation}
\begin{equation}
    \Sigma_\infty(R) = 2200 \text{ g/cm}^2\left(\frac{R}{1\text { AU}}\right)^{-3/2}.
\end{equation}
With this the density $\rho_\infty$ and $T_\infty$ are simply functions of $R$ as the characteristic background density may be estimated as $\rho_\infty \approx \Sigma_\infty/\sqrt{2\pi}H$ (in computing $H$ and in the remainder of this work we take the stellar mass $M_\ast = 1M_\odot$). Thus the number of free parameters is reduced by one and a steady state accretion model is well-defined simply given a planet mass, formation location, and luminosity $(M, R, L_\infty)$.

With only three parameters, it becomes feasible to compute a full grid of models spanning reasonable ranges in $\left(M, R, L_\infty\right)$. For a fiducial model grid, we computed $[40\times40\times40]$ logarithmically spaced points spanning $R\in\left[1, 300\right]$ AU and $M\in[20, 300] M_\oplus$. The span in luminosity is parameterized in terms of the adiabatic Bondi accretion rate $L_{\rm ad}\equiv GM\dot{M}_{\rm ad}/r_p$ with planet radius $r_p$. Defining the model luminosity as some fraction of the Bondi value $L_\infty\equiv b_L L_{\rm ad}$, the model grid spans $b_L\in[10^{-3}, 10]$. To estimate the planetary radius in calculation of $L_{\rm ad}$ we used the empirical cold giant radius relation of \citep{Thorngren+2019} as a suitable mass-radius relation $r_p(M)$, but we multiply this by a factor of 3 to parameterize the fact that these young accreting giants would be hotter and puffier than presently observed. For these fiducial models, we also assume dust opacities given by a dust-size distribution with power-law index $q=3.5$ and maximum grain size of $a_{\max}=1$ cm \citep{Zhu+2021}. Integrations of the steady-state equations were performed with $\texttt{LSODA}$ method of $\texttt{solve\_ivp}$ from the $\texttt{scipy}$ package and error tolerances $\texttt{rtol} = \texttt{atol} = 1.49012\times10^{-8}$. In all, we computed $64,000$ steady-state models with at least ten points per decade in each of $\left(M, R, L_\infty\right)$ in $\sim  4$ days or $\sim  5$ seconds per model.

With model grid in hand, and a choice of planet mass, orbital radius, and luminosity, one may calculate the steady-state accretion rate $\dot{M}(M,R,L_\infty)$. Until this point we have remained fairly agnostic about the emergent luminosity $L_\infty$. This is primarily due to the observational and theoretical considerations regarding an appropriate luminosity. Observationally, there simply are not bountiful accreting protoplanet detections at low mass or early time and corresponding luminosity constraints. Theoretically, one must know the intrinsic luminosity of the protoplanet as well as the properties of the accretion luminosity. Intrinsic luminosity estimates are the result of evolutionary models (e.g. \citet{MarleauCumming2014, Mordasini+2017}), subject to various model uncertainties. At the same time, accretion luminosity estimates are subject to the accretion mechanism (boundary-layer shock versus magnetospheric accretion) and uncertainties therein. 

In the absence of a single, more compelling model, it is worthwhile to consider applying our model grid to the simplest case of a boundary layer shock with shock efficiency $\eta_s$ at shock radius $r_s$ producing accretion luminosity $L_{\rm src}=\eta_s GM\dot{M}/r_s$ near the planetary surface. For the models presented in this work, we take all other sources of luminosity to be zero. This ignores, for example, the intrinsic luminosity of the young planet, which is sub-dominant for all but the lowest mass and slowly accreting planets \citep{Mordasini+2017, Marleau+2019}. To make sense of our model grid requires some way to associate the luminosity at the planetary surface with the escaping luminosity $L_{\rm src}\leftrightarrow L_\infty$. In principle, one could integrate from $L_\infty$ down to $r_s$ and measure $L_{\rm src}$. In Paper I however, we determined that luminosity tends to be nearly constant across the flow. We therefore simplify the computation by approximating the luminosity as constant and equating $L_{\rm src}\approx L_\infty$. Since luminosity acts as the coupling between flow inside the sonic point with flow outside the sonic point, this approximation makes integrating the interior solution entirely unnecessary and model grid integrations need only integrate up to the sonic point to determine the mass accretion rate. In Appendix \ref{sec:L} and the three dedicated models of \S\ref{sec:convect}, we test this approximation by doing the full integration to the planetary surface and getting the exact relation between $L_{\rm src}\leftrightarrow L_\infty$, but it makes minimal difference.

While the constant luminosity approximation simplifies the solution procedure by making $L_\infty = \eta_sGM\dot{M}/r_s$, there remains a complication in that the accretion rate itself depends on the luminosity $\dot{M}(M,R,L_\infty)$. Thus for each value of $M$ and $R$, as well as choices for $r_s$ and $\eta_s$, we have an implicit equation for $L_\infty\sim \dot{M}(L_\infty)$ which may be solved via standard root-finding across the $L_\infty$-dimension of our model grid. The found solution then is the one which has an accretion luminosity $L_\infty$ generated self-consistently by the steady-state $\dot{M}$. This procedure, by selecting out the self-consistent accretion $L_\infty$, reduces the dimensionality of the accretion rate parameter space from $\dot{M}(M,R,L_\infty)\rightarrow \dot{M}(M,R)$. In principle, this procedure could be done for many choices of $\eta_s$, $r_s$ and a transformed parameter space of $\dot{M}(M,R,\eta_s/r_s)$ constructed. Note that $\eta_s$ and $r_s$ are degenerate in this type of model, so really $\eta_s/r_s$ is the only relevant quantity. 

Here, we adopt a perfect shock efficiency $\eta_s=1$ assuming perfect conversion of potential energy to radiation. This choice is consistent with shock calculations \citep{Marleau+2017, Marleau+2019, ChenBai2022}, which generally find high efficiencies for $\gamma=7/5$. Even if the accretion shock is formally inefficient, from an energy conservation perspective, leftover potential energy gets radiated by a correspondingly hotter planetary envelope, making $\eta_s\approx 1$ appropriate in a global sense. For the shock radii, we take it equal to the empirical planet radii $r_p$ from cold giants measurements \citep{Thorngren+2019}, but again scaled up by a factor of three. An appropriate shock radius for these protoplanets remains an open question however (see \ref{sec:choice} for a different choice of $r_p$), with some models implying a significantly larger value given the planet's contraction may be quite prolonged \citep{GinzburgChiang2019-contraction}. 

\subsection{Fiducial Results}

\begin{figure*}
    \centering
    \includegraphics[width=\linewidth]{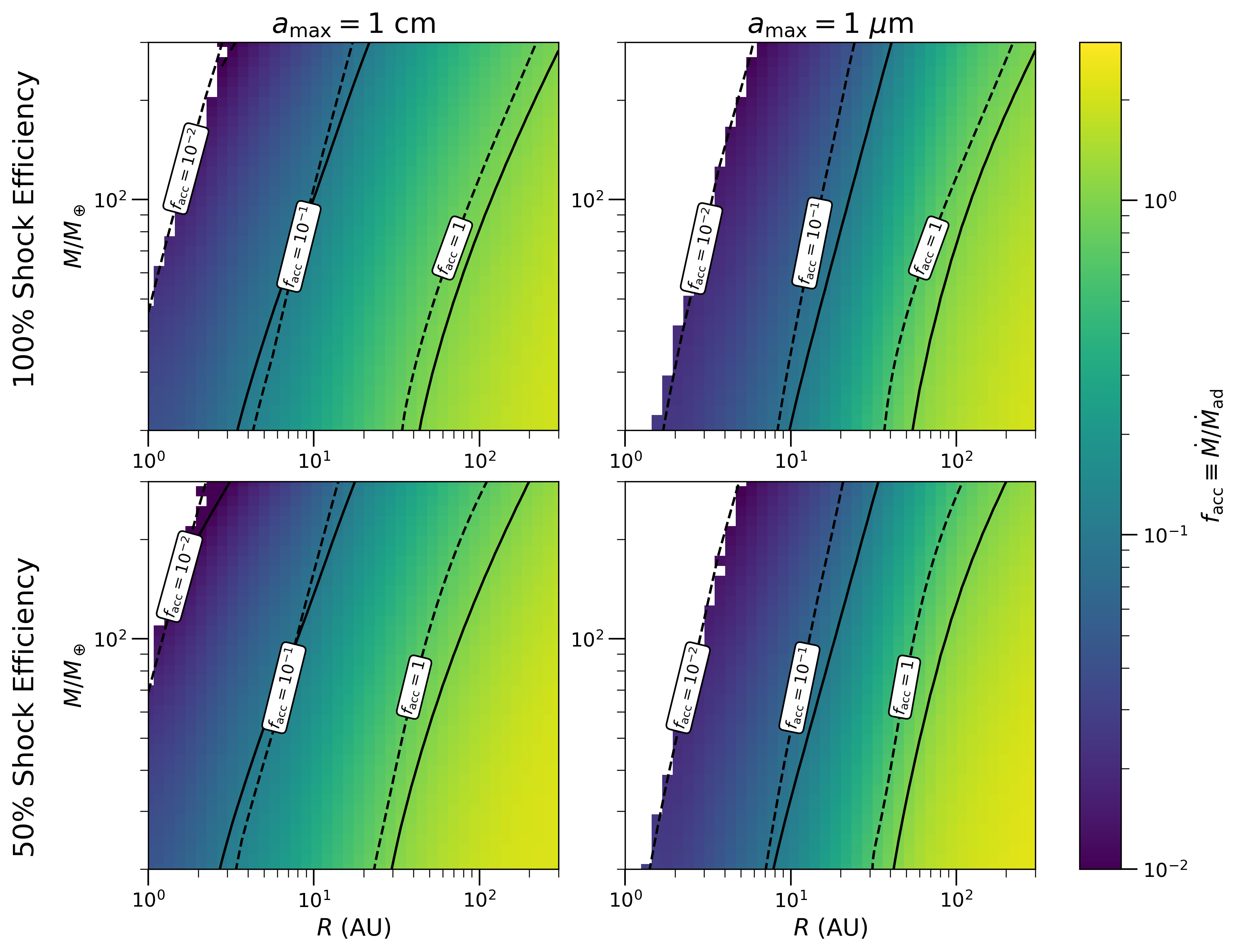}
    \caption{Steady-state accretion rates relative to the adiabatic Bondi rate as a function of planet mass and orbital distance. Left column shows models with dust opacities that adopt a maximum particle size $a_{\rm max}=1$ cm. Right column shows models from the grid testing a different opacity by changing the maximum particle size to $a_{\rm max}=10\mu$m. Upper rows select models with a luminosity assuming all accretion energy is converted to radiation at a planetary boundary layer shock. Lower rows assume this energy conversion is $50\%$ efficient (or a doubling of shock radius which is degenerate with efficiency). Regions where the model grid of this work found no steady-state transonic accretion solution are left blank. Contours of $f_{\rm acc}$ are also overlaid at the $[10^{-2}, 10^{-1}, 10^0]$ with solid lines. At the same levels, dashed contours estimating the accretion rates using the analytic formula (B7) from Paper I are shown for comparison.}
    \label{fig:fiducial}
\end{figure*}

With these choices, the parameter space of steady-state radiative accretion rates $\dot{M}(M,R)$ are shown in the upper-left panel of Figure \ref{fig:fiducial}. These accretion rates are shown normalized by the adiabatic Bondi rate $f_{\rm acc} = \dot{M}/\dot{M}_{\rm ad}$ such that they may be though of as a suppression or correction factor relative to adiabatic. In the outer disk, we see that $f_{\rm acc}\approx 1$, so that adiabatic rates are approximately correct. Throughout the inner disk ($\lesssim 10$ AU) however, we find a class of models where $f_{\rm acc} \lesssim 0.1$ and accretion is substantially suppressed by radiative feedback.

These results are consistent with the findings from our more general, dimensionless parameterization presented in paper I. We find that the accretion rate is well modeled with three dimensionless parameters that characterize how effectively radiation drives the accretion flow to be non-adiabatic. These are the characteristic optical depth $\tau_B\equiv\rho_\infty \kappa r_B$ through the Bondi radius $r_B=GM/(2c_{s,\infty}^2)$, the characteristic dimensionless luminosity $\tilde{L}_\infty \equiv L_\infty/(4\pi r_B^2 a_r c T_\infty^4)$, and a characteristic dimensionless cooling time
\begin{equation}
    \beta  \equiv  \frac{1}{4 \gamma (\gamma -1)} \frac{\rho_\infty c_{s,\infty}^3}{a c T_\infty^4} \frac{1}{\tau_B}
\end{equation}
With $\kappa\sim \kappa_R\sim 0.4$ cm$^2$/g $(T_\infty/100$ K)$^{1/2}$ suitable for our fiducial dust opacities, the characteristic optical depths and cooling times are simply,
\begin{equation}\label{eq:taub}
\tau_B = 0.5\left(\frac{M}{10 M_\oplus}\right)\left(\frac{R}{10 \text{ AU}}\right)^{-5/2} \left(\frac{M_\ast}{M_\odot}\right)^{1/2}
\end{equation}
\begin{equation}
\beta  = 0.04\left(\frac{M}{10 M_\oplus}\right)^{-1}\left(\frac{R}{10 \text{ AU}}\right)\ .
\end{equation}
The self-consistent dimensionless luminosity, on the other hand, varies little across this fiducial disk $\tilde{L}_\infty\approx 1-10$. In the outer disk ($\sim 100$ AU) where conditions are optically thin ($\tau_B\ll 1$, $\beta\lesssim 1$, $\tilde{L}\approx 5$), solutions fall into the nearly isothermal regime (see Paper I, Section 3.2.5). This corresponds to an accretion rate $f_{\rm acc}\approx 1$. Moving inwards through the protoplanetary disk, the accretion flow becomes optically thick and the solutions enter the ``Case 1" optically thick regime (see Paper I, Section 3.2.5). This regime of solutions is marked by a nearly constant luminosity profile and an accretion rate $f_{\rm acc}\sim (\tilde{L}_\infty\tau_B)^{-5/8}$. Because $\tilde{L}_\infty$ does not vary much across the disk and this scaling is $\beta$-independent, in this case, accretion suppression is governed by the optical depth and the trend of increasing optical depth with smaller orbital radius or higher mass seen in equation (\ref{eq:taub}). 

We verify this conclusion by taking the non-constant opacity models in Figure \ref{fig:fiducial} and for each $(M,R)$, computing $(\tau_B, \tilde{L}_\infty, \beta)$ with $\kappa =\kappa_R(\rho_\infty, T_\infty)$. In Paper I, we provide an analytic description of $f_{\rm acc}(\tau_B, \tilde{L}_\infty, \beta)$ (equation B4) which may be leveraged to estimate a corresponding constant opacity accretion rate for each non-constant opacity model here. Performing this mapping, we plot the contours of $f_{\rm acc}$ using the constant opacity rates in Figure \ref{fig:fiducial} and find good agreement (an average discrepancy of $26\%$ across the plotted parameter space) between non-constant opacity rates and the associated constant opacity ones. This suggests that the trend in $f_{\rm acc}$ is indeed driven by optical depth variations across the disk and that even constant opacity models are well-suited to reproduce this main trend with some fidelity.

In the very inner disk, at high mass (blank regions of Fig.\ \ref{fig:fiducial}), our initial value problem solver does not find a suitable steady-state solution. This is not entirely surprising because these models are in the super-thermal regime $(r_B> H)$ and the location of the sonic point $\sim r_B$ becomes comparable to the location of the outer boundary $\sim H$ in these models. Indeed, the constant opacity framework of Paper I returns steady-state solutions in this regime, but in that case the accretion flow is treated as having infinite extent ($H\rightarrow \infty$). We conclude that while perhaps this regime has viable Bondi-like solutions in an idealized (Paper I) sense, the lack of solutions here is physically motivated more than anything.

\subsection{Choices of Dust Opacity, Planet Radius, and Shock Efficiency}\label{sec:choice}
Having constructed a reasonable fiducial model demonstrating that high optical depth in the inner disk suppresses planetary growth rates, we begin to relax some of the assumptions of the fiducial model and test the robustness of the results. Since the emergent accretion suppression manifests predominantly from an opacity effect, let us first examine a separate opacity prescription assuming a different but reasonable dust distribution. Instead of a dust distribution with maximum particle size 1 cm, we now consider a distribution with only small grains and $a_{\max} = 10$ $\mu$m, but still using the compiled giant planet opacity tables of \citet{Zhu+2021}. In this case, below sublimation temperatures, the opacities approximately scale as $\kappa\sim T^{2}$ akin to widely used icy regime of \citet{BellLin1994}. Performing the same solution procedure as in the fiducial models, we arrive at the upper right panel of Figure \ref{fig:fiducial}. The general effect is to decrease $f_{\rm acc}$, i.e.\ compress the contours of $f_{\rm acc}$ and/or shift them to larger $R$, though this effect appears to be fairly modest only modifying rates by  factors of $\approx 1-3$. 

Next we examine the effect of our choice of planetary radius and thus the location of the shock radius $r_s$ where accretion energy is released in this model. In the fiducial model we artificially inflated the radii of present giant planets \citep{Thorngren+2019} by a factor of three to describe our accreting planet radii. Here we test radii artificially inflated by an additional factor of two. Because the planet radius and the shock efficiency are degenerate parameters in $L\sim\eta_s GM\dot{M}/r_s$, this could equivalently be thought of as a test of lowering the shock efficiency from $100\%\rightarrow 50 \% $. Applying this larger radius/lower shock efficiency to the model grid for both opacity distributions, we arrive at the lower two panels in Figure \ref{fig:fiducial}. Comparing the accretion rates to the previous models, the effect of the factor of two increase in shock radius is fairly mild. It strictly acts to increase the accretion rate with the resulting increase about $\sim10-30\%$ in magnitude with some minor dependence on $(M,R)$. Taking all these tests together, the observed fiducial $f_{\rm acc}$ appears to be fairly robust to changes in the assumed dust distribution, planet radius, and shock efficiency.

\subsection{The Effect of Gap-Opening}
While the preceding modifications to the fiducial models were relatively small in magnitude, the process of gap-opening may be substantially more impactful. The disk surface density can drop by orders of magnitude in gaps,  driving a proportionally large reduction in the key parameter determining $f_{\rm acc}$, the optical depth. It is prudent then, to test the robustness of the derived accretion rates under the effects of gap-opening and to identify regimes where gap-opening is expected to operate.

\begin{figure}
    \centering
    \includegraphics[width=\linewidth]{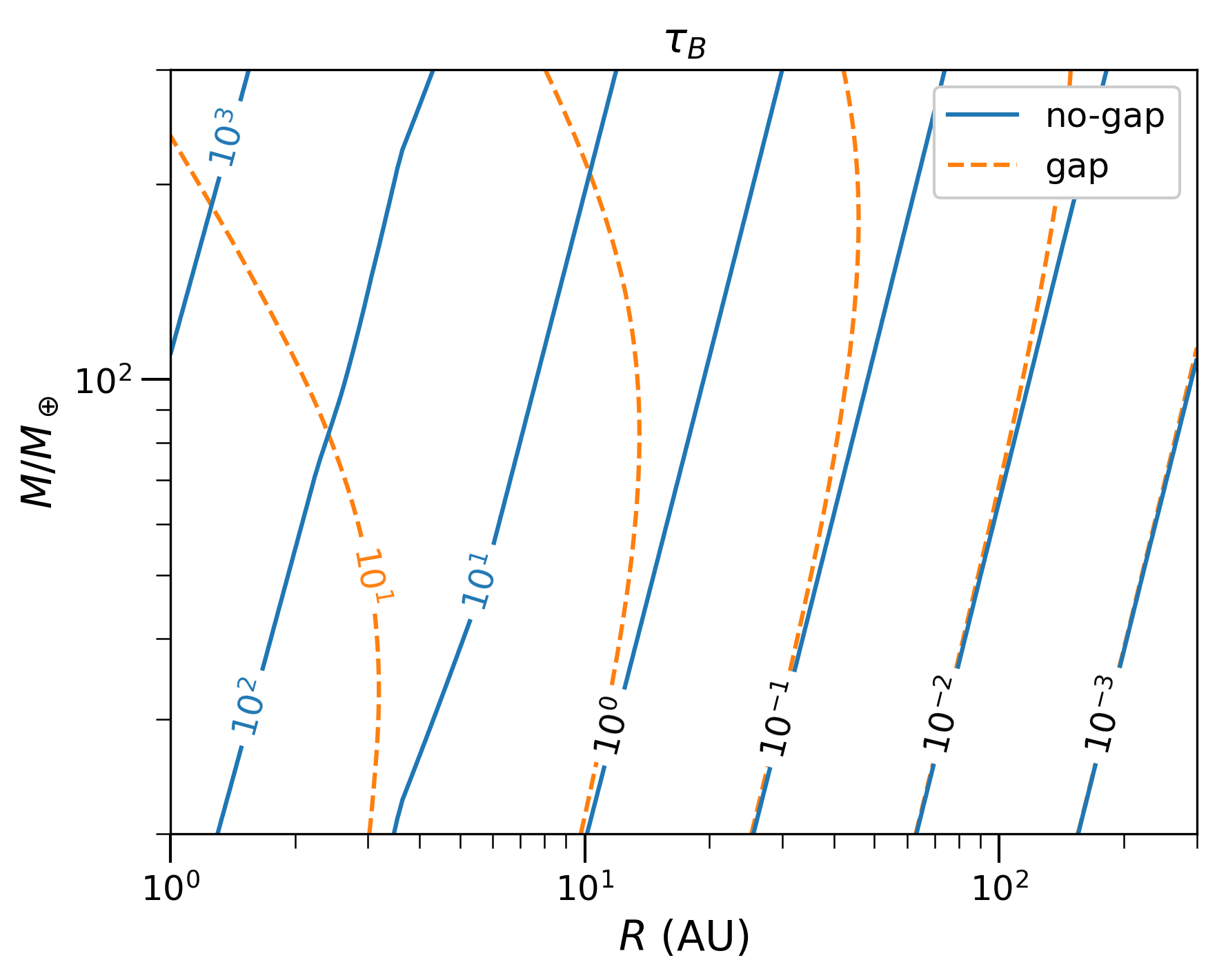}
    \caption{Contours of the characteristic optical depth $\tau_B$ ($\sim \rho_\infty$) across the fiducial parameter space for gap and no-gap models. Their ratio is the gap depth  $D$.}
    \label{fig:taub}
\end{figure}

To do so, we modify our fiducial disk model $\Sigma(R)$ to now include a parameterization of planet carved gaps \citep{DuffellMacFadyen2013, Fung+2014, Kanagawa+2015}, $\Sigma_{\rm gap} = \Sigma(R)D(M/M_\ast,H/R,\alpha)$, requiring introduction of a new free parameter -- the standard disk $\alpha$-viscosity. In this context, we take $\alpha=2\times 10^{-3}$ consistent with the standard adoption in planet population synthesis models \citep{BurnMordasini2024}. The gap depth $D(M,R,\alpha)$ is taken as the prescription given in \citet{Kanagawa+2016}, $D(M,H/R,\alpha)\equiv (1+0.04K)^{-1}$ with $K\equiv (M/M_\ast)^2(H/R)^{-5}\alpha^{-1}$. The remainder of the procedure for computing 1D steady-state models remains the same as the fiducial case and the role of gap-opening is in modifying the characteristic density $\rho_\infty$ and therefore $\tau_B$. In Figure \ref{fig:taub}, we give an estimate of the magnitude of the gap depths by plotting $\tau_B$ (also a proxy for $\rho_\infty)$ in both the fiducial no-gap and gap cases. Exterior to $\sim 10$ AU, gaps are shallow and the effect on these models is negligible. In the inner disk however, densities begin to diverge with greater discrepancies at large planet mass to potentially great effect. 

\begin{figure}
    \centering
    \includegraphics[width=\linewidth]{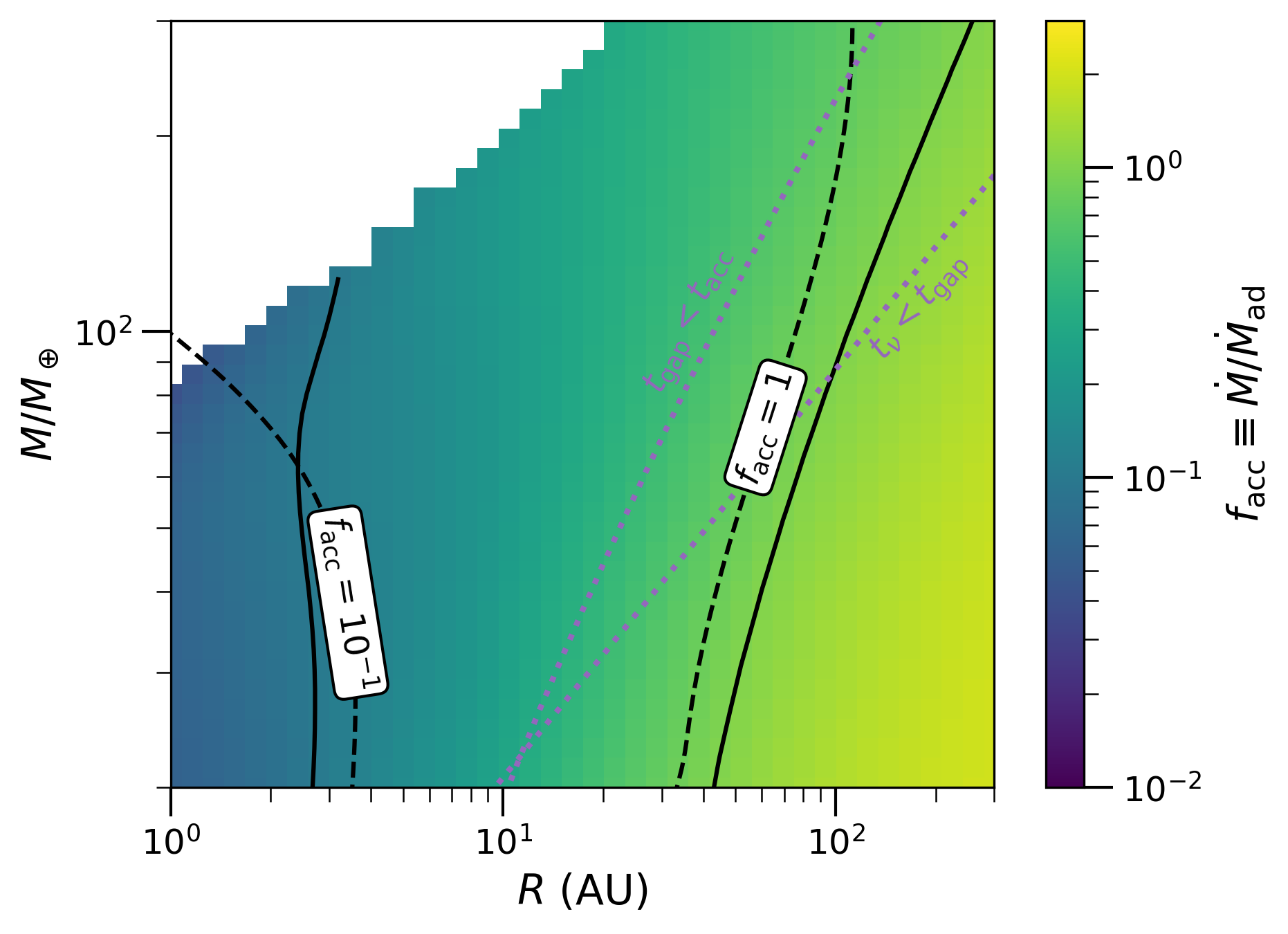}
    \caption{Mass accretion rates in the style of Figure \ref{fig:fiducial}, but for models incorporating the density perturbation induced by a gap. Dotted lines are drawn where $t_{\rm gap} = t_{\rm acc}$ and $t_{\nu} = t_{\rm gap}$.}
    \label{fig:gap}
\end{figure}

To quantify this effect, we repeat the calculation of steady-state accretion rates across the fiducial model parameter space but with a nebular density set by the steady-state gap depth with $\alpha=2\times 10^{-3}$. The accretion rates of these ``gapped" models are presented in Figure \ref{fig:gap}. Towards low mass, where planets open shallow gaps, the results are relatively unchanged, and an order-of-magnitude suppression in the accretion rate at $1-5$ AU persists. At high mass, gaps in the inner disk are made substantially deep and more adiabatic accretion rates are expected. In practice, where these greatest gapped/non-gapped discrepancies are expected, is also a regime where solutions are not found because $r_B > H$ and Bondi accretion is not physically applicable. However, the constant opacity models from Paper I, in which the boundary is placed at infinity still provide estimates of $f_{\rm acc}$. These estimates (dashed contours in Figure \ref{fig:gap}) indeed suggest a much weaker suppression at high mass than implied by our fiducial non-gapped models. Whereas in the non-gapped models of Figure \ref{fig:fiducial}, $f_{\rm acc}\approx 10^{-2}$ at high mass, in Figure \ref{fig:gap}, there exist no contours extending to $f_{\rm acc}=10^{-2}$ and the constant-opacity dashed contours suggest $f_{\rm acc}$ at high mass lies in the range of $f_{\rm acc}\approx 0.1-1$.

\subsubsection{The Conditions for Gap-Opening}
This begs the question: which accretion rates are expected -- gapped or non-gapped? Consider the relevant gap-opening $t_{\rm gap}$ and accretion timescales $t_{\rm acc}\equiv M/\dot{M}$. When $t_{\rm gap} < t_{\rm acc}$, gap-opening is the dominant process and gapped accretion rates in the manner of Figure \ref{fig:gap} are more appropriate. For a gap of width $\Delta R$, a gap-opening timescale is found by considering the angular momentum difference required to transport gas across the gap:
\begin{equation}
l \sim \Sigma \Omega (R\Delta R)^2
\end{equation} and the rate at which angular momentum can be transferred -- that is, the angular momentum flux carried away by density-waves \citep{GoldreichTremaine1980, Rafikov2002}
\begin{equation}
\dot{l}\sim \frac{(GM)^2\Sigma R\Omega}{c_s^3}.
\end{equation} 
Thus a gap-opening timescale is given by
\begin{equation}
    t_{\rm gap}\sim \frac{l}{\dot{l}}\sim \Omega^{-1}q_t^{-2}\left(\frac{H}{R}\right) \left(\frac{\Delta R}{H}\right)^2
\end{equation}
 where $q_t \equiv (M/M_\ast)(H/R)^{-3}$ is the planet's dimensionless thermal mass. Since simulations \citep{DongFung2017} have measured gap-widths $\sim H$ and our outer boundary for the steady-state models was also taken at $H$, taking $\Delta R\sim H$ is suitable for the simple estimates of gap-opening here. This gives
 \begin{equation}
     \frac{t_{\rm gap}}{t_{\rm acc}} \sim\left(\frac{M_{\rm disk}}{M}\right)f_{\rm acc}
 \end{equation}
 with $M_{\rm disk}\equiv \Sigma RH$, approximating the enclosed disk mass.

We plot the estimated line for $t_{\rm gap}=t_{\rm acc}$ in Figure \ref{fig:gap} with models left $(t_{\rm gap }<t_{\rm acc})$ of this line more appropriately adopting gapped accretion rates. Gap-opening in a viscous disk however also requires that the gap-opening action operate more strongly than the viscous action to fill the gap. Thus to open a gap, we also require $t_{\rm gap}< t_\nu\equiv \Delta R^2/\nu$. The line $t_{\rm gap}=t_\nu$ is also shown in Figure \ref{fig:gap}, with models left of this line readily able to open a gap even in the face of viscous diffusion. From this we conclude that essentially all models with $t_{\rm gap} < t_{\rm acc}$ also have $t_{\rm gap}<t_\nu$, and are therefore prone to gap-opening. Because the parameter space in which gapped and non-gapped models are discrepant are likely to have gaps, we recommend the gap-opened models of Figure \ref{fig:gap} as being more representative of planetary accretion rates in general.

For the purposes of implementing radiative feedback into population synthesis models, this timescale analysis is immaterial. The steady-state accretion rates are simply a function of optical depth which is readily estimated from the disk surface density returned each timestep of an evolutionary calculation. This highlights the utility of coupling our radiative feedback model to time-dependent calculations (see \S \ref{sec:tools}).

\section{Time-Dependent Simulations}\label{sec:convect}
\subsection{Comparison with Steady-State Solutions}
To verify the results of Section \ref{sec:fiducial}, several models were chosen to simulate in 1D and run to steady-state in Athena++ as outlined in Section \ref{sec:time}. In particular, we ran three models corresponding to a $50M_\oplus$ planet at 2, 10, and 100 AU. For direct comparison, we also compute their steady-state solution by solving the initial value problem (IVP) of equations (\ref{eq:rho})-(\ref{eq:er}), integrating down to the planetary radius (as opposed to \S\ref{sec:models} which only integrates to the sonic point). Here a $2$ AU model was preferred over $1$ AU because near the planetary surface of that model, the accretion flow reaches $10^4$ K, ionizing Hydrogen and abruptly increasing the opacity by four orders of magnitude. While this would not effect the accretion rate, (occurring interior to the sonic point) this increases the resolution requirement substantially, so instead a $2$ AU model (peak temperature $3500$ K) was simulated. 

Each simulation domain spans $[r_p,H]$ and uses the same $a_{\rm max} = 1$ cm opacity table as the fiducial steady-state models. The corresponding physical parameters as well as other relevant quantities are listed for reference in Table \ref{tab:sim}. Each simulation adopted a root grid of $128$ logarithmically spaced cells and resolves the subsonic flow region with two additional levels of mesh refinement to increase resolution without increasing the timestep. Meanwhile, the angular resolution of the radiative transfer was set to $N_\theta=8$ polar angles spanning the $(0,2\pi)$ range. This was increased to $N_\theta =16$ angles for the $R=100$ AU model, necessitated by the optically thinness of the gas at such large $R$. Each simulation was run to $t_{\rm sim}=10(H/c_{s,\infty})$ corresponding to at least ten sound-crossing times -- sufficient to relax to steady-state. 

\begin{figure*}
    \centering
    \includegraphics[width=\linewidth]{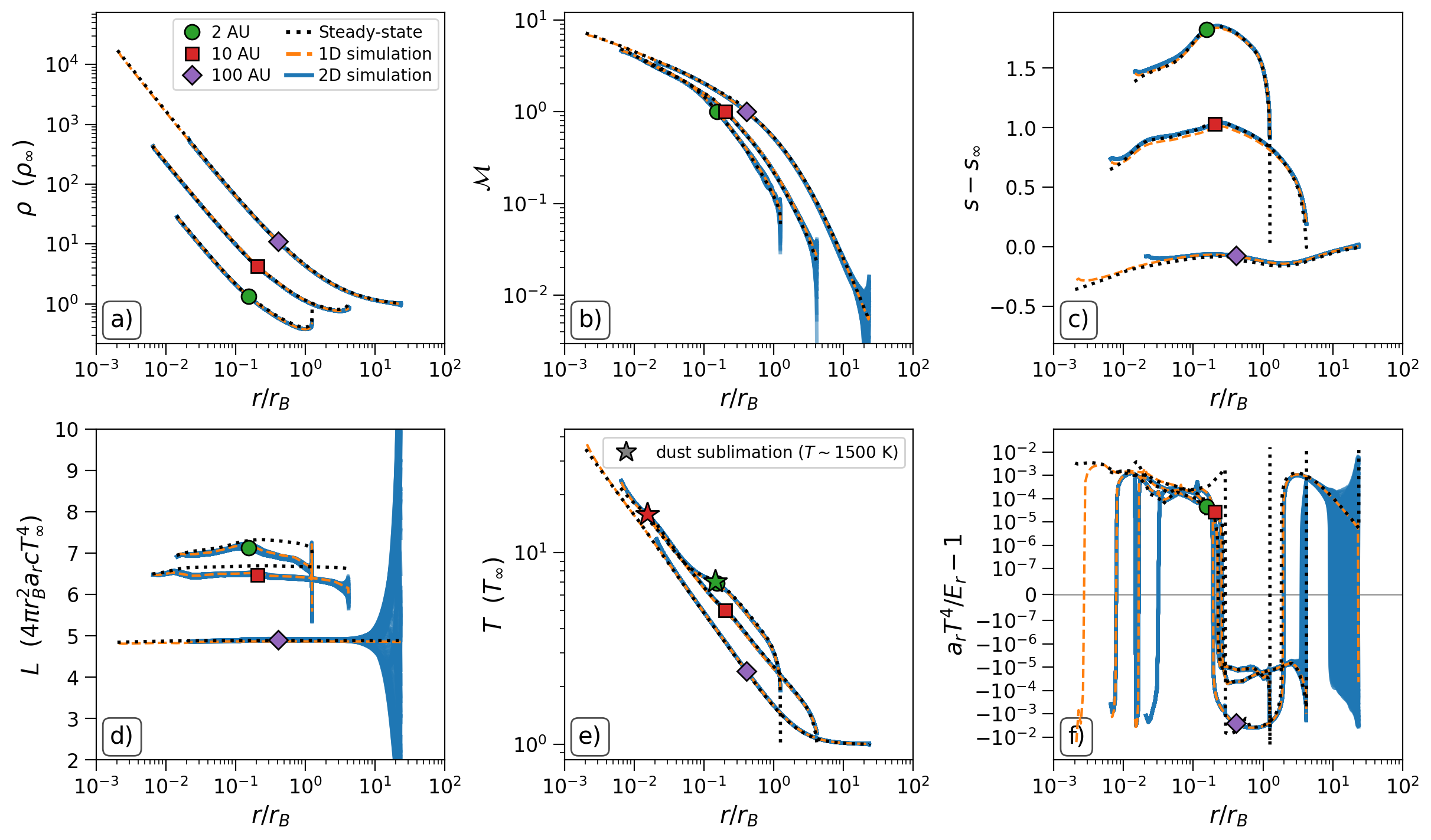}
    \caption{Comparison of flow solutions for a $50M_\oplus$ planet at $2$, 10, and 100 AU, between our steady-state IVP solve (dotted) versus 1D (dashed) \& 2D (solid) Athena++ simulations. Each panel is a different flow variable: density, Mach number, entropy, luminosity, temperature, and fractional disequilibrium. Markers (circle, square, diamond) denote the location of the sonic point. Stars mark the radius at which dust sublimation occurs.}
    \label{fig:sim}
\end{figure*}

The resulting 1D flow profiles at simulation end are plotted in Figure \ref{fig:sim} and compared against steady-state solutions of the IVP. Overall we find remarkable concurrence between the time-dependent and steady-state solutions across all three simulations. These simulations each trace physically distinct parameter space: The 100 AU model is optically thin everywhere -- there is no photosphere anywhere and radiation free streams from $r_p$ to $H$. The 10 AU model is a marginally thin/thick $\tau_B\approx 1$ case. Near the hot planetary surface, dust is sublimated and opacity is very low. The dust sublimation radius occurs at $\sim$ a few $r_p$ but the flow remains optically thin until nearing the sonic point. Beyond that scale, up to $H$, it stays optically thick. The 2 AU model is formally very optically thick $\tau_B\approx 100$. However, the flow is much hotter and dust sublimation occurs near the sonic point where opacity changes by 1-2 orders of magnitude. The result is a structure which is entirely optically thin inside the sonic point and optically thick outside. 
In Paper I, it was predicted that optically thin gas in the free-falling interior should have positive entropy gradients. It is not surprising then that is the case in panel c) because all three models here have optically thin interiors.

Panel f) shows the level of disequilibrium between the radiation and gas $a_rT^4/E_r-1$. We find that all the models are close to radiative equilibrium across the entire flow, as expected in planet-forming regimes where $\beta\ll 1$. The disequilibrium profiles tend to diverge in sign at the inner boundary between the IVP solutions and simulations. This is because our simple simulation boundary condition (outflow+constant luminosity), does not enforce or extrapolate the appropriate level of disequilibrium into the ghost cells. %Obviously
This mismatch does not effect the results since perturbation to the hydrodynamic variables is negligible.

Similarly, the luminosity profiles show marginal disagreement between simulations and IVP. This is because the code does not conserve total energy with radiation (so $ds/dr\sim dL/dr$ is not enforced) and the relative error in the luminosity perturbation ends up greater than any of the other variables. We have performed convergence testing doubling the resolution up to 4096 cells for the 10 AU case and found that the luminosity does linearly converge to match the steady-state profile (L1-norm: $5.934\%\rightarrow 4.039\%\rightarrow 2.145\%\rightarrow 1.0729\%\rightarrow 0.5304\%$). Though again, being under-converged in the luminosity perturbation at the level of several per-cent does not change the accretion rate since the error in the hydrodynamic variables is substantially lower. The steady-state solutions for 2 and 10 AU do show physical luminosity variation at the discernable level of $\sim 5\%$, but this is not significant enough to break our constant luminosity approximation of \S\ref{sec:models} (see also Appendix \ref{sec:L}).

\begin{deluxetable}{llll}
%\digitalasset
\tablewidth{0pt}
\tablecaption{1D Simulation Parameters \label{tab:sim}}
\tablehead{
\colhead{Value} & \colhead{2 AU} & \colhead{10 AU} & \colhead{100 AU}
}
\startdata
$\rho_\infty$ (g/cm$^3$) & $2.5\times 10^{-10}$ & $3.0\times 10^{-12}$ & $5.4\times 10^{-15}$\\
$\gamma$ & $7/5$ & (-) & (-) \\
$\mu$ & 2.4 & (-) & (-) \\
$T_\infty$ (K) & 212 & 95 & 30\\
$M$ ($M_\oplus$) & $50$ & (-) & (-) \\
$L_\infty$ (erg/s) & $3.8\times 10^{31}$ & $7.2\times 10^{30}$ & $5.3\times 10^{29}$ \\
$L_{\rm src}$ (erg/s) & $3.8\times 10^{31}$ & $7.0\times 10^{30}$ & $5.2\times 10^{29}$ \\
$r_p/r_B$ & $1.4\times 10^{-2}$ & $6.4\times 10^{-3}$ & $2.0\times 10^{-3}$\\
$H/r_B$ & $1.3$ & $4.2$ & 24\\
$\tau_B$ & $91$ & $2.6$ & $7.7\times 10^{-3}$\\
$\beta$ & $2.8\times 10^{-3}$ & $9.0\times 10^{-3}$ & $9.5\times 10^{-2}$\\
$f_{\rm acc}$ & $0.034$ & $0.161$ & $1.20$\\
%$\kappa_R(r_0)$ (cm$^2$/g) & & & \\
%$\kappa_P(r_0)$ (cm$^2$/g)
\enddata
\tablecomments{Entries with (-) take the preceding left value in the same row. $L_{\rm src}$, $L_\infty$ values are evaluated from the non-constant $L$ solutions of the steady-state equations in Figure \ref{fig:sim}.}
\end{deluxetable}

\subsection{Convection}
In Paper I and preceding works \citep{Flammang1984, Markovic1995}, the computed entropy profiles indicate that these accretion flows are potentially unstable to convection. As evidenced in Figure \ref{fig:sim}, the outer solutions of models with radiatively suppressed accretion also show inverted (negative) entropy gradients. In the large $r$ limit, the flow velocity goes to zero and the situation is approximately in hydrostatic equilibrium, meeting the conditions for Schwarzchild instability. Because of this, it is these outermost regions where convection is most favored. In Paper I, however, it was seen that upon nearing the Bondi radius, many models with convective exteriors are prone to developing a radiative-convective boundary (RCB) and transitioning to a radiative region which persists up to the sonic point. Depending on the influence of this radiative buffer zone and the efficiency of convection, the role of convective energy transport may be limited in these planetary contexts.

\begin{figure*}
    \centering
    \includegraphics[width=\linewidth]{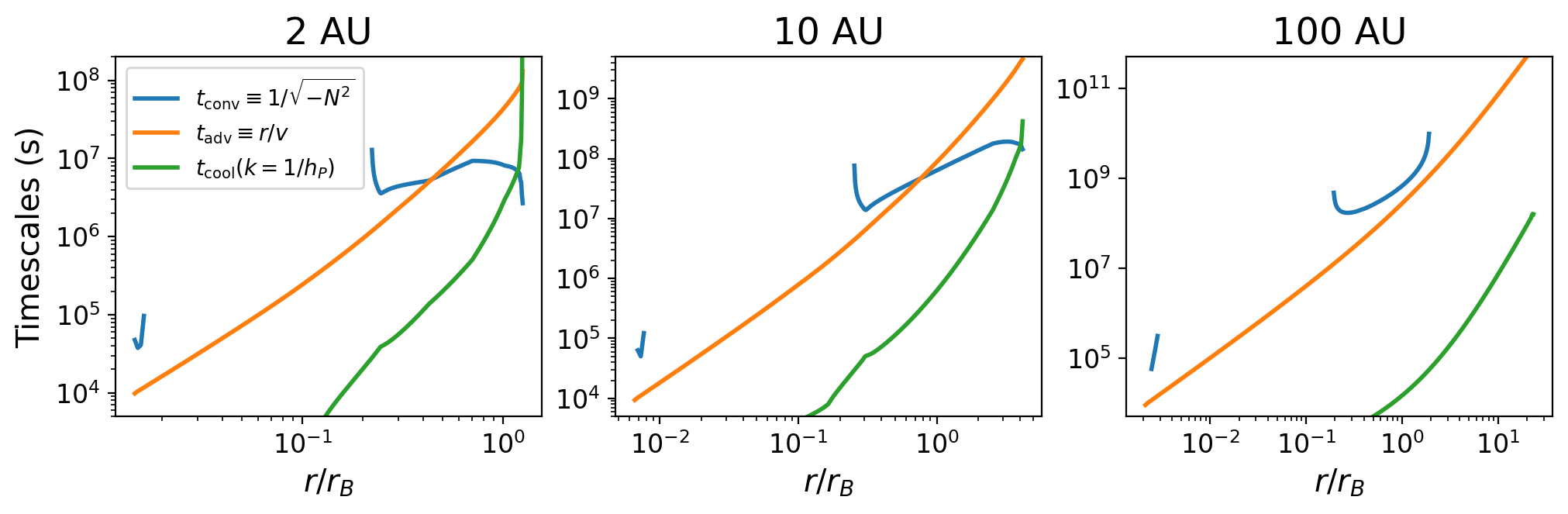}
    \caption{Radial variation of timescales associated with each 1D simulation. }
    \label{fig:scales}
\end{figure*}

To estimate whether convection is expected to operate in the $r<H$ extent of the models here we compare the relevant timescales in the 1D simulation profiles. The timescales of interest are the flow/advective/dynamical timescale $t_{\rm adv}\equiv r/v$, the convective timescale -- the inverse Brunt Va\"{i}sala frequency, $t_{\rm conv}\equiv 1/\sqrt{-N^2}$, and the characteristic radiative cooling time of a perturbation with wavenumber $k$ \citep{UnnoSpiegel1966, MihalasMihalas1984} 
\begin{equation}
    t_{\rm cool}^{-1}\equiv   \frac{4a_rc\kappa_P T^3}{c_P}\left(\frac{1}{1 + 3\rho^2\kappa_R\kappa_P/k^2}\right).
\end{equation}
In this convective context, a suitable wavenumber is posited to be something like an inverse pressure scale height $h_P^{-1}\equiv (d\ln P/dr)$. If $t_{\rm adv}< t_{\rm conv}$, convective elements are sheared apart by background flow before becoming unstable. Likewise, if $t_{\rm cool}< t_{\rm conv}$ convective elements radiatively cool and bulk fluid motion becomes an inefficient mechanism for energy transport. Thus a suitable but approximate condition for convection to develop is $t_{\rm conv}<t_{\rm adv}, t_{\rm cool}$. By comparing these timescales in Figure \ref{fig:scales}, it is demonstrated that cooling is very rapid in all simulations and the character of the flow should be radiative in nature. There is a very small region at the outer boundary where the convective time is technically small enough to be convective in the 2 AU model, but we suspect that this is not entirely physical because calculation of the cooling time with $k\sim h_P^{-1}$ is incorrect when the convective layer itself is $\ll h_P$ (see following discussion and accompanying figure). Even if convection developed there it would be small and inefficient, flattening the local (steep) entropy gradient at the boundary but not changing the overall character of the flow. 

\begin{figure}
    \centering
    \includegraphics[width=\linewidth]{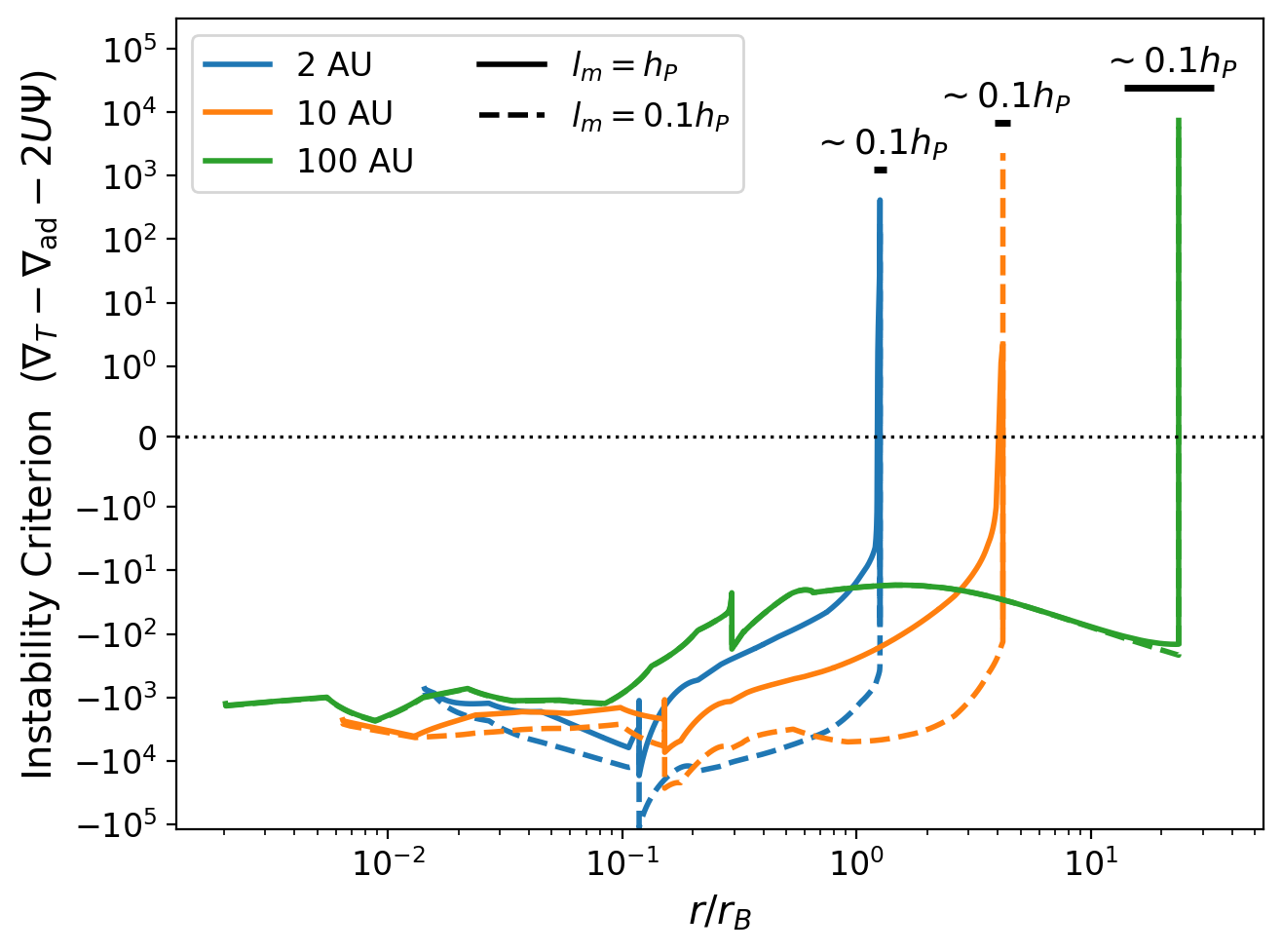}
    \caption{Instability criterion evaluated for the steady-state models at 2, 10, 100 AU. Positive values are formally unstable. Solid lines choose a mixing length $l_m=h_P$, while dashed lines choose a more appropriate $l_m=0.1h_P$. The approximate scale of $0.1h_P$ at the location of the ``unstable'' region is also shown.
    \label{fig:unstable}}
\end{figure}

These 1D profiles also fail to develop convection under the mixing length formalism of \citet{Markovic1995}. In this case, the requirement for convection is modified from the Schwarzchild criterion $\nabla_T - \nabla_{\rm ad} > 0$ to a stricter $\nabla_T - \nabla_{\rm ad} > 2U \Psi$, with $U$ quantifying the action of heat fluxes and $\Psi$ the radial shear (see Paper I for details). Choosing a mixing length $l_m=h_p$ and computing this modified instability condition on the steady-state solutions gives the solid curves presented in Figure \ref{fig:unstable}. As suggested by the timescale arguments, the rapid cooling time results in very inefficient convection -- so inefficient that the background radial flow stabilizes convection throughout nearly the entirety of the flow. The exception is a unstable spike occurring just near the outer boundary where the entropy gradient is steepest. This presumes however that the mixing length is $h_p$ when the width of the unstable spike is in fact substantially smaller. If one recomputes the instability criterion using a mixing length $l_m=0.1h_P$, closer to the width of the unstable region, the unstable region becomes negligibly small in width. The remaining ``unstable'' region in this case is certainly unphysical as it persists even at the outer boundary of the $100$ AU model where the flow is not expected to be convective (see middle panel of Figure \ref{fig:sim}). As there is no self-consistent unstable region of the flow under this mixing length formalism, we conclude the absence of convection in the 2D simulations is physical and the recovered steady-state entropy profiles robust. 

To put the convective character of our 1D solutions to the true test however, we also extended the 1D radial simulations along an additional angular dimension $\theta$ (co-latitude) with periodic boundary conditions $[-\pi/4, \pi/4]$ and ran 2D time-dependent simulations keeping most of the setup unchanged. While the character of convection is very different between 2D and 3D, testing the stability and some qualitative measure of the efficiency of convective elements is reasonably explored by running computationally cheaper 2D simulations. In going to 2D, we augment the polar angles of the radiative transfer grid with $8$ angles along each azimuth. Thus each cell calculates the radiative transfer along $64$ rays in the $2,10$ AU simulations and $128$ rays in the $100$ AU simulation. To improve the cost for the most expensive 100 AU simulation, we artificially move the inner boundary outwards by a factor of ten and adjust the refinement grid to keep the radial width of each cell unchanged between 1D/2D. Each of the 2, 10, 100 AU simulations has 40, 32, 48 cells on each root grid respectively, to keep the cells approximately square. With these choices, the simulations resolve $\sim 50$ cells per pressure scale height near the sonic point, increasing to $\sim 100$ at the outer boundary. We break the symmetry of the simulations by continuously forcing perturbations at the outer boundary. These are implemented by augmenting the density in the ghost cells with solutions for plane-like waves of wavenumber $k_\theta =k_r h_P\equiv k_n$ propagating with speed $c_{s,\infty}$:
\begin{equation}
    \rho = \rho_\infty  + A\sum_n \sin\left(k_n\theta + \phi_{n}\right)\sin\left(\frac{k_nc_{s,\infty}}{H}t + \psi_n\right)
\end{equation}
We seed six wavemodes ($n=1,2,4,8,16,32$) with randomly distributed phases $\phi_n$, $\psi_n$, amplitude $A=0.02\rho_\infty$, and wavenumbers $k_n = 2\pi n/ (\theta_{\max} - \theta_{\min}) = 2\pi n/(\pi/2)$, such that the physical wavelength of each mode is $\sim (1.6, 0.8, 0.4, 0.2, 0.1, 0.05)$ scale-heights respectively.

Ultimately, none of the 2D simulations show any evidence of convective instability or energy transport to alter the steady-state entropy profiles. This can be seen in Figure \ref{fig:sim}, where also plotted are the profiles along each angle for the final snapshot in each 2D simulation. Again, the 2D profiles show remarkable agreement with the steady-state solution with some additional scatter induced by the perturbative forcing at the outer boundary, but no sign of convection or alteration to the mean entropy. 

\section{Open-Source Tools}\label{sec:tools}
To incorporate the effects of radiative feedback into existing/future evolutionary calculations and population synthesis codes, we develop a lightweight, flexible set of open-source tools to compute $f_{\rm acc}$ for given planet/disk parameters. In Section \ref{sec:models}, it was demonstrated that the constant opacity models of $f_{\rm acc}$ in Paper I reasonably approximate the $f_{\rm acc}$ of more realistic models presented here. In the interest of flexibility, we therefore implement the constant opacity parameterizations of $f_{\rm acc}$ -- both through analytic equation (B4) and tabulated numerical solutions $f_{\rm acc}(\tau_B,\tilde{L}_\infty, \beta)$ of Paper I. We provide a repository\footref{repo} containing the raw tabulated ASCII data $f_{\rm acc}(\tau_B,\tilde{L}_\infty, \beta)$ from Paper I, simple Python functions and wrappers to evaluate $f_{\rm acc}$ (\texttt{bonditools.py}), as well as a \texttt{Jupyter} notebook of usage examples reproducing figures from this series of papers and demonstrating how to merge these tools with population synthesis outputs.

We supply two functions in \texttt{bonditools.py} intended as user-endpoints for evaluating $f_{\rm acc}$. The first, \texttt{compute\_facc\_from\_cgs}, calculates the suppression factor $f_{\rm acc}(M, \rho_\infty, T_\infty,\gamma,\mu, L_\infty, \kappa)$ given a set planet/disk parameters in centimeter-gram-second (CGS) units. Since we use constant opacity solutions to parameterize $f_{\rm acc}$, users should supply a single opacity characteristic to their problem -- e.g.\ $\kappa \approx \kappa_R(\rho_\infty, T_\infty)$ in optically thick or $\kappa \approx  \kappa_P(\rho_\infty, T_\infty)$ in optically thin. This function requires some knowledge of the luminosity $L_\infty$, which, depending on the problem, may or may not be known a priori. For this reason, a second function, \texttt{compute\_luminosity\_facc\_from\_cgs}, exists to evaluate both a self-consistent accretion luminosity and $f_{\rm acc}$ simultaneously. This function evaluates $f_{\rm acc}$ subject to the constraint,
\begin{equation}
    L_\infty  = \frac{\eta_s GM f_{\rm acc}\dot{M}_{\rm ad}}{r_s} + L_{\rm other}
\end{equation}
where the first term is a self-consistent accretion luminosity $L_{\rm acc}$ and $L_{\rm other}$ includes any other desired luminosity sources which do not directly depend on $f_{\rm acc}$ (e.g.\ intrinsic planetary luminosity). Thus the function evaluates $f_{\rm acc}(M, \rho_\infty, T_\infty,\gamma,\mu, \eta_s, r_s, L_{\rm other}, \kappa)$, where all quantities are known at each timestep, $(\eta_s\sim 1, r_s\sim r_p)$ of a standard evolutionary calculation.

Our recommended usage, and thus the default settings (\texttt{mode="mixed"}), are to use the analytic formula in regimes where the underlying assumptions are satisfied ($\beta \leq 1$) and interpolation/extrapolation of the tabulated numerical solutions elsewhere. For reasonable planet forming conditions, $\beta \lesssim 1$, reducing to the analytic parameterization. This choice may be explicitly overwritten by setting the function argument \texttt{mode="tabulated"}, along with other various numerical preferences. 

These functions are designed to be readily imported into existing population synthesis and evolutionary calculations. For codes which do not use Python or wrapping of these functions into the calculation is difficult, one may use these functions to pre-compute a lookup table of $f_{\rm acc}(\beta, \eta_s/r_s, \tau_B)$ for example, and then simply evaluate these parameters and lookup $f_{\rm acc}$ at each timestep of the evolution. Since our functions are vectorized and there are only three\footnote{four, if one includes $L_{\rm other}\neq 0$} free parameters, this is straightforwardly and easily done (see \texttt{examples.ipynb}). 

\section{Conclusion}\label{sec:end}
We have demonstrated that radiative feedback during the accretion process substantially decreases the accretion rates of giant planets in the runaway growth phase. In these models, optical depth through the Bondi radius is the primary determinant of the efficacy of radiative feedback. Consequently this effect is negligible in the outer ($>10$ AU) disk, whereas accretion rates for planets at $1$ AU can be suppressed by up to two orders of magnitude. Because massive planets in the inner disk are also more prone to gap-opening, which reduces the optical depth, we estimate that suppression is typically $\sim 1$ order-of-magnitude in the inner disk, biased toward lower $\lesssim 100 M_\oplus$ planet masses where  gaps are shallower.

We highlight several important consequences of reduced accretion rates found here. First, our models imply lower luminosities (by $f_{\rm acc}$) for accreting protoplanets, potentially reducing their detectability in future surveys. Secondly, by extending the growth time of planets by up to an order of magnitude in some regimes, the shape of the mass function of close-in planets could be impacted. In particular, this effect might smooth out the predicted bi-modal distribution from classic core accretion models \citep{IdaLin2004,Mordasini+2009}. Implementation of the modified accretion rates in population synthesis models are necessary to establish these trends more clearly, as the final impact may depend strongly on the fraction of a planet's accretion history in which it is embedded in a gap. See \S \ref{sec:tools} and our open-source tools\footref{repo} for simple methods to incorporate these effects into existing calculations.

%% Please use the acknowledgment and contribution environments. This will 
%% be anonomyized when the "anonymous" style option is used. 
\begin{acknowledgments}
We thank Zhaohuan Zhu for useful discussions. This work is supported by the National Aeronautics and Space Administration under Agreement No. 80NSSC21K0593 for the program
“Alien Earths.” This work has also been supported by National Aeronautics and Space Administration under Agreement No. 80NSSC24K0163. This material is based upon High Performance Computing (HPC) resources supported by the University of Arizona TRIF, UITS, and Research, Innovation, and Impact (RII) and maintained by the UArizona Research Technologies department. Resources supporting this work were also provided by the NASA High-End Computing (HEC) Program through the NASA Advanced Supercomputing (NAS) Division at Ames Research Center.
\end{acknowledgments}

\section*{Data Availability Statement}
Tabulated data for $f_{\rm acc}$ and other tools are available at https://github.com/apbailey/radiative-bondi-products. Additional data underlying this article will be shared on reasonable request to the corresponding author.

\begin{contribution}
AB led the production of this work -- development of numerical solutions, analysis of the models, and writing of the manuscript. KK \& AY provided substantial guidance and feedback on the manuscript, models, and analysis.
\end{contribution}

%\facilities{HST(STIS), Swift(XRT and UVOT), AAVSO, CTIO:1.3m, CTIO:1.5m, CXO}

\software{This work made extensive use of the SciPy \citep{2020SciPy}, Matplotlib \citep{Matplotlib}, and NumPy \citep{Numpy} packages. Radiation-hydrodynamics simulations were performed with the publicly available version of the Athena++ code \citep{Stone+2020, Jiang2021}. Claude Sonnet 4.6 \citep{anthropic2026claude} was used to perform minor optimizations and improve readability for \texttt{bonditools.py}}


\bibliography{sample7}{}
\bibliographystyle{aasjournalv7}



\appendix

\section{Self-Gravity} \label{sec:selfg}
\begin{figure}
    \centering
    \includegraphics[width=0.5\linewidth]{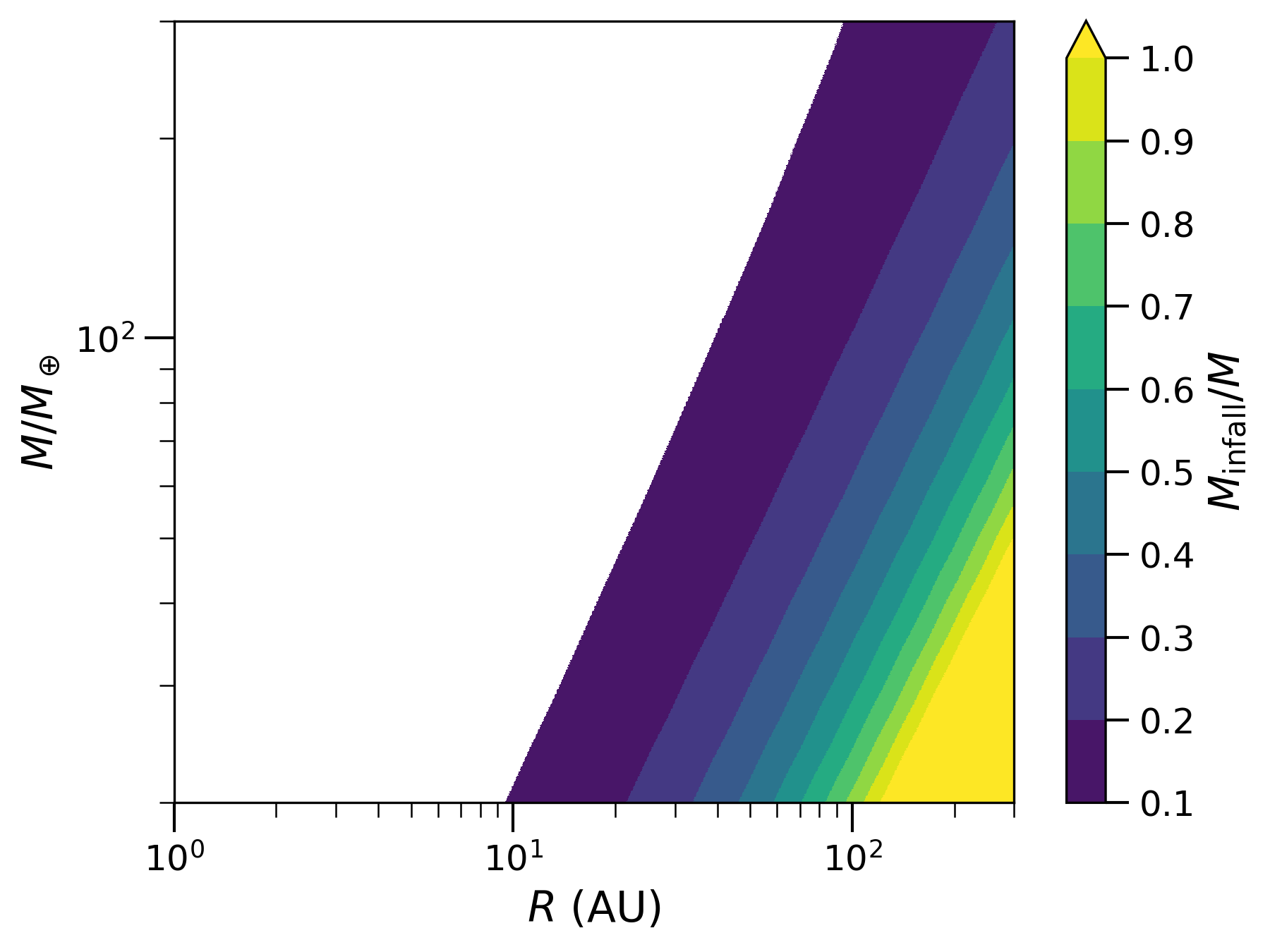}
    \caption{The ratio of mass contained in an adiabatic $\gamma=7/5$ accretion flow ${M}_{\rm infall}$ to the assumed central planetary mass $M$ for our fiducial disk model.}
    \label{fig:selfg}
\end{figure}

In this work, the gravity source is treated as a Newtonian potential representative of a central protoplanet, while any self-gravitating contribution from the accretion flow itself has been neglected. This appendix serves to estimate the parameter space where this self-gravitating contribution may be significant and the magnitude of the effect this has on the results presented. Unfortunately, the potential contributed by the accretion flow itself is not outright calculable from the models presented in Section \ref{sec:models} since the interior solution is obfuscated by our inability to integrate through the sonic point within the current formalism. Nevertheless, as a reasonable estimate, one can calculate the mass of an adiabatic Bondi flow ${M}_{\rm infall}$ for planets in the fiducial disk model with their corresponding inner/outer flow boundaries $[r_p, H]$. This integrated adiabatic mass is plotted relative to the central planet mass assumed in this work in Figure \ref{fig:selfg}.
We see that in much, but not all, of the parameter space investigated, the mass of the flow is only a fraction of the central mass and neglect of self-gravity is reasonable. For planets at $R\gtrsim100$ AU with mass $M \lesssim 50M_\oplus$ however, the mass contained in the adiabatic accretion flow exceeds the mass of the central planet.

\section{Validity of the Constant Luminosity Approximation}\label{sec:L}
In the model grid of \S\ref{sec:models} and the construction of Figs.\ \ref{fig:fiducial}, \ref{fig:gap} the luminosity was treated as being nearly constant along the entire flow. This follows from the results of Paper I, where the luminosity is constrained to be nearly constant down to the shock surface $r_s$ so long as the flow is near freefall. In \S\ref{sec:models}, this approximation made it possible to use the luminosity at the \textit{outer} boundary $L_\infty \approx L_{\rm src} = \eta_sGM\dot{M}/r_s$ as the consistency condition for solving for the mass accretion rate. When done this way, the interior solution is irrelevant in determining the mass accretion rate, removing the need for integration of the interior solution and simplifying the calculations. 

Here we examine the validity of the $L_{\rm src}\approx L_\infty$ approximation and magnitude of the effect on our derived accretion rates. To do so, we calculate $f_{\rm acc}$ for a set of steady-state models using the truly correct $L_{\rm src} = \eta_sGM\dot{M}/r_s$ consistency condition. This requires, that for each planet mass and location $(M, R)$, solutions are integrated down to the planetary radius to evaluate the value of $L_{\rm src}$. Furthermore, the value of $L_\infty$ is iterated upon and solutions re-integrated until $L_{\rm src} = \eta_sGM\dot{M}/r_s$ is satisfied. For this calculation we take $\eta_s=1$ and $a_{\rm max}=1$ cm, as in the upper-left panel of Fig.\ \ref{fig:fiducial}. 

\begin{figure}
    \centering
    \includegraphics[width=\linewidth]{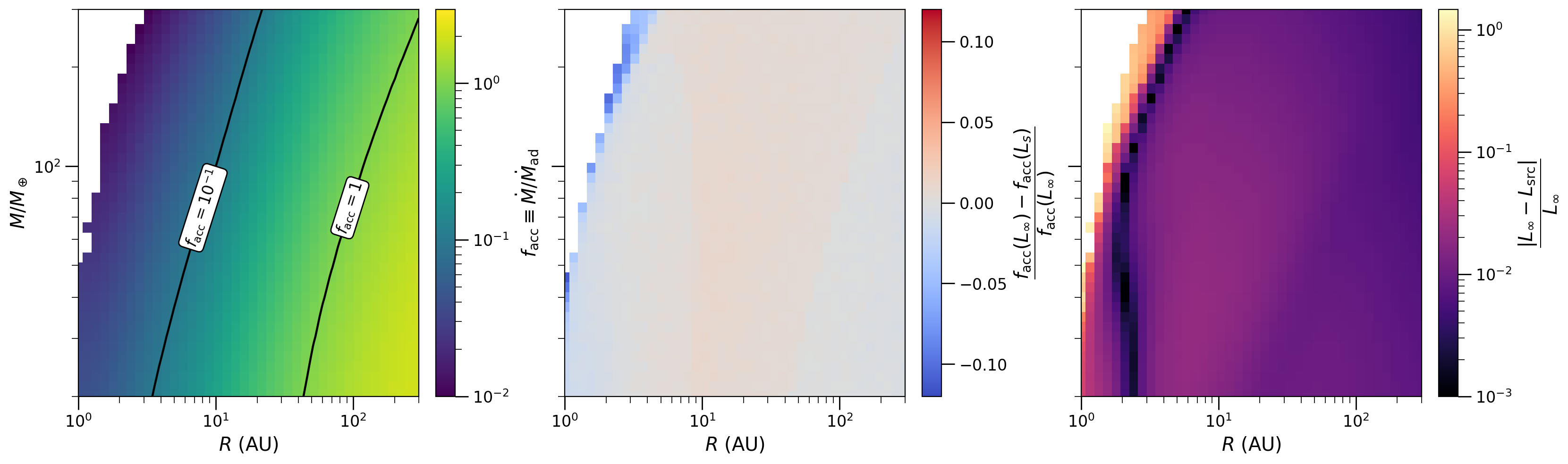}
    \caption{Left: $f_{\rm acc}$ computed using the correct luminosity at the inner boundary $L_{\rm src}$ instead of the outer boundary $L_\infty$. Center: the fractional correction to the resulting accretion rate $f_{\rm acc}$ when compared with the results of \S\ref{sec:fiducial}. Right: the fractional difference in interior/exterior luminosity in integrations extending to planetary surface.}
    \label{fig:L}
\end{figure}

The results of this calculation using the interior luminosity are shown in Fig.\ \ref{fig:L}. The left panel shows the derived $f_{\rm acc}$; to be compared against the upper-left panel of Fig.\ 1. The middle panel shows the fractional change in derived $f_{\rm acc}$ whether one uses the correct luminosity $L_{\rm src}$ or $L_\infty$ (the constant luminosity assumption). We find the correction to be at the per-cent level ($\sim$ the level of our iteration tolerance), reaching a maximum of $12\%$ at the inner edge of the disk. The right panel takes the computed luminosity profile in this calculation and estimates the change across the profile by comparing the fractional change between the inner and outer boundary. Again, this difference is at the per-cent level except for the line of cells straddling the innermost $R$. We therefore conclude that the correction obtained by integrating the interior solution and using the interior accretion luminosity is minimal and the outer integrations/results of \S\ref{sec:models} are suitable. 

\end{document}